\documentclass[preprints,article,accept,pdftex,moreauthors]{Definitions/mdpi} 

\firstpage{1} 
\pubvolume{1}
\issuenum{1}
\articlenumber{0}
\pubyear{2024}
\copyrightyear{2024}
\datereceived{ } 
\daterevised{ } 
\dateaccepted{ } 
\datepublished{ } 
\hreflink{https://doi.org/} 

\def\beqn{\begin{eqnarray}}
\def\eeqn{\end{eqnarray}}
\def\beq{\begin{equation}}
\def\eeq{\end{equation}}

\newcommand{\cc}[2]{c{#1\atopwithdelims()#2}}

\Title{Spinor--Vector Duality and Mirror Symmetry}

\TitleCitation{SVDMS}

\Author{Alon E. Faraggi $^{1}$,*\orcidA{0000-0001-7123-6414}}

\AuthorNames{Alon E. Faraggi}

\AuthorCitation{Faraggi, A.E.}

\address{%
$^{1}$ \quad Mathematical Sciences Department, University of Liverpool, Liverpool L69 7ZL, United Kingdom}

\corres{alon.faraggi@liv.ac.uk}

\abstract{\\
  {\it
Mirror symmetry was first observed in worldsheet string constructions
and shown to have profound implications in the Effective Field
Theory (EFT) limit of string compactifications, and for the properties of
Calabi-Yau manifolds.
It opened up a new field in pure mathematics and was utilised in the
area of enumerative geometry.
Spinor-Vector Duality (SVD) is an extension of mirror symmetry.
This can be readily understood in terms of the moduli of toroidal
compactification of the heterotic string, which include the metric
the antisymmetric tensor field and the Wilson line moduli.
In terms of toroidal moduli, mirror symmetry corresponds to mappings
of the internal space moduli, whereas spinor-vector duality corresponds
to maps of the Wilson line moduli.
In the past few of years, we demonstrated the existence of spinor-vector duality in the effective field theory compactifications of the string theories. This was achieved by starting with a worldsheet orbifold construction that exhibited spinor-vector duality and resolving the orbifold singularities,
hence generating a smooth effective field theory limit with an imprint of
the spinor-vector duality. Just like mirror symmetry,
the spinor-vector duality can be used to study the properties of complex
manifolds with vector bundles.
Spinor--vector duality offers a top--down approach to
the ``Swampland''--program, by exploring the imprint of the
symmetries of the ultra--violet complete worldsheet string
constructions in the effective field theory limit. The SVD suggests
a demarcation line between (2,0) EFTs that possess an ultra--violet
complete embedding versus those that do not. 
  }\\
~~\\
LTH--1379}


\begin{document}




\section{Introduction}

Physics is first and foremost an experimental science. Be that as
it may, the language which is used to encode the experimental data is
mathematics. We build mathematical models that aim to describe the
experimental data. A successful mathematical model is one that is
able to account for the wider range of observational data.

On the other hand, continued advances in the experimental probes of
matter require new mathematical tools to account for the experimental
observations. The objective of describing wider ranges of experimental
results using common mathematical frameworks leads to new insight
into the required mathematical structures. 

Experimental data in the sub--atomic domain since the end of the 19th century,
cumulated in the Standard Model of particle physics.
This model utilises the framework of relativistic 
point Quantum Field Theories (QFTs) to account for a wide
range of experimental observations in the sub--atomic realm.
The Standard Model consists of three sectors and
is underlined by invariance
of its basic equations under spacetime and internal transformations.
The gauge interactions in the Standard Model are
mediated by spacetime vector bosons that transform
as spin 1 representations of the Poincare group. 
The gauge sector contains three group factors that account for the
strong, weak and electromagnetic interactions.
The matter sector of the Standard Model consists of three
generations of chiral spin 1/2 states that carry identical charges
under the three gauge sectors of the Standard Model.
Finally, the Higgs sector of the Standard Model consists
of a single spin 0 electroweak doublet that facilitates
the synthesis of the short range weak interactions
with the long range electromagnetic interactions.
Ellis, Nanopoulos and Gaillard were among the first to
advocate the experimental searches for the Standard Model
Higgs boson \cite{EGN1976}.

The success of the Standard Model opened the door to Grand Unified
Theories (GUTs) \cite{PS1974, GG1974, GQW1974, BEGN1977} in which the
three gauge groups of the Standard Model are unified into
one simple GUT group and the matter and scalar sectors appear
in representations of the GUT group. The development of GUTs
is a watershed in the progression of physics because the perceived
unification can only be manifest at a scale which is far removed
from energy scales that are currently probed by experiments. 
Grand Unified Theories and this vast separation are supported by several
experimental observations. In particular, the multiplet structure of the
Standard Model states that are embedded in representations of the GUT group.
In the context of $SO(10)$ GUTs \cite{SO10GUTS},
embedding of the Standard Model matter
states entails that the number of free parameters that are required
to account for the matter states gauge charges is reduced from 54
to one. A remarkable coincidence indeed. Additionally, the vast
separation between the GUT scale and the electroweak scale is
supported by the longevity of the proton; the suppression of
left--handed neutrino masses; and the logarithmic evolution of the Standard
Model parameters, which is compatible with the obervational data in the
gauge sector and the heavy generation matter sector.

The high GUT scale, however, introduces a problem. While the lightness of the
gauge and matter sectors compared to the GUT scale can be explained by the
existence of symmetries that protect them from being drawn to the GUT scale,
nothing protects the scalar sector from this fate. To explain the lightness
of the scalar sector we can use a new spacetime symmetry, supersymmetry, or
assume that the scalar states transform under a new gauge sector, which
become strongly interacting near the electroweak scale. These proposals
will be tested in future collider experiments.

Grand Unified Theories and their supersymmetric extention introduce
a new twist in the tale. If supersymmetry is localised it requires the
inclusion of a spin 2 state, which is the mediator of the
gravitational interactions, in the theory. Localised supersymmetry
forces the merger of the gauge and gravitational interactions.
The road, however, does not end there. Gravity is incosistent
as a point quantum field theory. It is plugged with infinities.
It is then extremely rewarding that a small departure from
point quantum field theories provides a consistent framework for
the synthesis of the gauge and gravitational interactions. The
local supersymmetric extensions of the Standard Model
are effective field theory limits of string theory,
in which the point--like idealisation of elementary
particles is replaced by a string representation.

String theory provides an elaborate mathematical structure that
unifies the gauge, matter and scalar sectors of the Standard Model
with gravity. Its self--consistency conditions imply that additional
degrees of freedom, beyond those that are observed in the Standard Model,
should exist in nature. Some of these degrees of freedom may
be interpreted as additional spacetime dimensions that are
made small enough to evade detection by contemporary experiments,
whereas others may be interpreted as additional gauge symmetries,
beyond those in the Standard Model.
The extra spacetime degrees of freedom can be compactified on an
internal six dimensional real manifold or a three dimensional
complex manifold.
The compactified spaces determine many of the
properties of the observed Standard Model matter
spectrum, like the number of chiral generations and their
masses.
In this manner string theory gives rise to models
that reproduce the main phenomenological properties of the
Standard Model.
In particular, string theory provides a framework in which
the Yukawa couplings of the Standard Model fermionic states
to the scalar Higgs can be calculated in terms of the
gauge coupling.
This is a remarkable feature of string theory that provides a
framework to calculate the Standard Model fermion masses.

The self--consistency conditions of string theory require
the introduction of new worldsheet degrees of freedom that in some
guise may be interpreted as extra spacetime dimensions. In ten
dimensions we have five supersymmetric theories that together
with 11 dimensional supergravity are believed to be effective perturbative
limits of a more fundamental theory that is traditionally dubbed M--theory.
Additionally, string theory in ten dimension gives rise to a tachyon free
non--supersymmetric vacuum and seven non--supersymmetric vacua that
are tachyonic and unstable.
The heterotic $E_8\times E_8$ string is the effective
stable string theory limit that reproduces the GUT picture hinted at by
the Standard Model data, as it is the only string theory that gives
rise to spinorial representations in the perturbative spectrum.

Phenomenological string models that reproduce the main phenomenological
properties of the Standard Model, {\it i.e.} three chiral generations and
the correct charges under the Standard Model gauge group, were
constructed since the mid--eighties. The free fermionic formulation of the
heterotic--string in four dimensions \cite{ABK, KLT, AB}
led to a particular class of quasi--realistic worldsheet contructions.
These models correspond to toroidal $Z_2\times Z_2$ orbifolds of six
dimensional compactified tori at special points in the moduli space
\cite{FFFZ2Z2, AFGNM}.
The fermionic $Z_2\times Z_2$ orbifolds provide a laboratory
to explore how the detailed phenomenology of the
Standard Model and unification emerge from string theory.
Among there are:
\begin{itemize}
\item construction of the first Minimal Standard Heterotic--String Model
  (MSHSM) that contains solely the states of the Minimal Supersymmetric
  Standard Model (MSSM) in the effective low energy field theory
  below the string scale \cite{FNY, CFN}.
\item The prediction of the top quark mass at $\sim 175-180$GeV \cite{topqmp}. 
\item Fermion masses and CKM mixing \cite{CKM}. 
\item Neutrino masses \cite{FHNuMasses}.
\item Gauge coupling unification \cite{gaugecu}.
\item Proton stability \cite{protonstability}.
\item Supersymmetry breaking \cite{FKP}. 
\item Moduli fixing \cite{moduli}.
  
\end{itemize}

The quasi--realistic free fermionic models motivate a deeper
investigation of this class of string compactifications.
Specifically, they highlight the potential relevance of
the structure of the toroidal $Z_2\times Z_2$ orbifolds.
A few remarks are in order here. The first is in regard to
the existence of a string landscape. The number of string vacua
in ten dimensions is relatively small. Five supersymmetric and
eight non--supersymmetric. However, in four dimensions
the number of vacua is enormous, with some authors quoting the
number $10^{500}$ or even more. The meaning of this expansive space is
yet to be understood. One theme alluded to in this paper is that
they may all in fact be connected. Our task is to unravel and to
understand the symmetries that underlie this vast space of possibilities
and their inner connections. However, even if this space is enormous
it is still believed to be finite. In each one of these possibilities
the parameters are supposedly determined in terms of the Vacuum
Expectation Value (VEV) of a few fixed moduli, that determine the
characteristics of the internally compactified space. This should
be contrasted with the Standard Model (SM) that contains 19
continuous parameters, {\it i.e} an infinite 19 dimensional
space, and its Beyond the Standard Model (BSM) QFT extensions
that contain numerous more continuous parameters.
String constructions are constrained by the straitjacket of
quantum gravity. Although the space of possibilities is
vast, it is finite rather than infinite, in contrast to point
particle QFT contructions that are not constrained by
the consistency conditions imposed by quantum gravity.
Furthermore, despite the fact that the space of string vacua is vast,
the majority of string compactifications are not directly relevant to
the observable world, {\it i.e.} they contain too many moduli fields and
too many chiral generations to allow for a viable connection with the
observable parameters of the Standard Model. To date the fermionic
$Z_2\times Z_2$ orbifolds have been studied in most detail and
provide concrete quasi--realistic examples to study how
the parameters of the Standard Model can be determined
in a theory of quantum gravity. Since their early days
they provide a framework to explore the phenomenology of the Standard
Model on the one hand and the mathematical properties that
underlie string theory on the other hand. Spinor--vector Duality (SVD)
and its relation to mirror symmetry are among these mathematical
properties.

\section{Spinor--Vector Duality}\label{svd}

While the early free fermionic models 
consisted of isolated examples \cite{revamp, FNY, alr, slm, slmcon}, since
2003 systematic computerised methods were developed that enable scans
of large number of free fermionic heterotic--string vacua with different
unbroken $SO(10)$ subgroups \cite{fknr, fkr, acfkr, frs, slmclass, asyclass}.
Similar computerised tools for the classification of type II superstring
were developed in \cite{GKR}. 
A recent comprehensive review of this subject is provided in
\cite{FRReview}.
The computerised classifications tools facilitated the discovery of a
remarkable symmetry that underlies the space of (2,0) heterotic--string
compactifications, which is akin to mirror symmetry. Spinor--Vector
Duality operates in vacua in which the $N=4$ spacetime supersymmetry is
broken from $N=4$ to $N=2$ or $N=1$ by a $Z_2$ or $Z_2\times Z_2$,
respectively, of the internal compactified coordinates. The gauge
symmetry from the ten dimensional $E_8\times E_8$ symmetry
then depends on the action of a Wilson line. In the absence
of a Wilson line the gauge symmetries are $E_8\times E_8$ in the $N=4$
case, $E_7\times SU(2)\times E_8$ in the $N=2$ case, and
$E_6\times U(1)^2\times E_8$ in the $N=1$ case.
The twists of the internal coordinates produce
twisted sectors that give rise to massless states in 
the $56$ representation of $E_7$, and $27$ and $\overline{27}$
representation of $E_6$. 
The inclusion
of a specific Wilson line breaks the gauge symmetries in the three cases
to $SO(16)\times SO(16)$, $SO(12)\times SU(2)\times SU(2)\times SO(16)$
and $SO(10)\times U(1)^3\times SO(16)$, respectively.
In terms of the unbroken subgroups the $56$, $27$ and $\overline{27}$
are decomposed into spinorial and vectorial representations
of the unbroken subgroup, which are $(32,1)$, $(32^\prime,1)$ and
$(12, 2)$ in the case of $SO(12)\times SU(2)$, and
$16$, $\overline{16}$ and $10$ in the case of $SO(10)\times U(1)$. 
The Spinor--Vector Duality (SVD) operates under the exchange
of the spinor and vector representations.
Focusing on the $N=1$ case, the $27$ and $\overline{27}$
representations of $E_6$ decompose as $27=16_{+1/2}+10_{-1}+1_{+2}$
and $\overline{27}=\overline{16}_{-1/2}+10_{+1}+1_{-2}$
under $SO(10)\times U(1)$. It is seen that in the case of
$E_6$ for every spinorial $16$ state there is a vectorial $10$
state and for every $\overline{16}$ there is a vectorial $10$
state. Vacua with $E_6$ symmetry are therefore self--dual
if we exchange the total number of $16\oplus\overline{16}$
representations of $SO(10)$ with the total number of
vectorial $10$ representations. The remarkable property is
that there is a remnat of this symmetry when the $E_6$
symmetry is broken to $SO(10)\times U(1)$. The statement
is that for a string vacuum with a number $\#_1(16+\overline{16})$
of spinorial and anti--spinorial $SO(10)$ representations,
and a $\#_2(10)$ of vectorial representations, there is a
dual vacuum in which the two numbers are interchanged, {\it i.e.}
$\#_1\leftrightarrow\#_2$ \cite{FKRSVD, FKRNeq2, CJFKR, AFT}.

The Spinor--Vector Duality resembles $T$--duality in the sense
that the enhanced symmetry point with $E_6$ symmetry is
self--dual under the SVD. The duality can also be seen to
operate in terms of a spectral flow operator 
on the bosonic side of the heterotic--string \cite{FKRNeq2, FFMT}. 
The vacua that possess $E_6$ symmetry have $(2,2)$ worldsheet
supersymmetry. On the supersymmetric side of the heterotic--string,
a vector in the basis that defines the free fermionic
models operates as a spectral flow operator that mix between
sectors that produce spacetime fermions and bosons. Similarly,
on the bosonic side of the heterotic--string, in the case with
enhanced $E_6$ symmetry and $(2,2)$ worldsheet supersymmetry,
the models can be constructed such that the operation of
a spectral flow operator on the bosonic side is manifest
\cite{FKRNeq2, FFMT}.
In this case, the spectral flow operator on the bosonic side
mixes between the spinorial and vectorial representations
in the breaking of $E_6$ under $SO(10)\times U(1)$.
In the vacua in which the $E_6$ symmetry is broken to $SO(10)\times U(1)$
and the $N=2$ worldsheet supersymmetry on the bosonic side is
broken, the spectral flow operator induces the map between the dual
vacua.
\begin{table}[H] 
\setlength{\tabcolsep}{5mm}
\begin{tabularx}{\textwidth}{ccc|ccc|ccc|c}
\toprule
 \multicolumn{3}{c}{\textbf{First Plane}} &  \multicolumn{3}{c}{ \textbf{Second Plane}} &    \multicolumn{3}{c}{\textbf{Third Plane }}&  \\
\midrule
\boldmath{$s$} &\boldmath{${\bar s}$}& \boldmath{$v$} &\boldmath{$s$}&\boldmath{${\bar s}$}&\boldmath{$v$}&\boldmath{$s$}&\boldmath{${\bar s}$}&\boldmath{$v$}& \textbf{\# of Models} \\
\midrule
2 & 0 & 0 &    0 & 0 & 0 &    0 & 0 & 0 & 1325963712 \\
0 & 2 & 0 &    0 & 0 & 0 &    0 & 0 & 0 & 1340075584 \\
1 & 1 & 0 &    0 & 0 & 0 &    0 & 0 & 0 & 3718991872 \\
\midrule
0 & 0 & 2 &    0 & 0 & 0 &    0 & 0 & 0 & 6385031168 \\
\bottomrule
\end{tabularx}
\caption{Number of models with a total number of 2 representations in
  the first twisted sector. 
}
\label{svdcounting}
\end{table}

\begin{figure}[h]
\includegraphics[width=14pc]{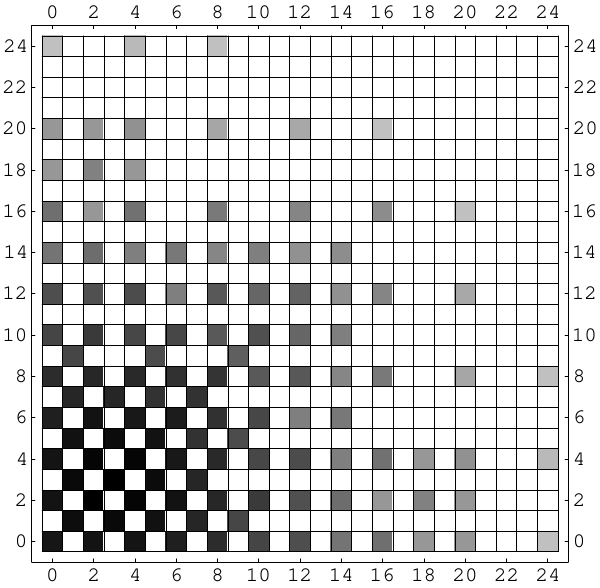}\hspace{0.5pc}%
\begin{minipage}[b]{20pc}
\caption{
\label{den}
Density plot depicting the spinor--vector duality in the space of fermionic
$Z_2\times Z_2$ heterotic--string orbifolds. The figure shows the number 
of models with a given number of $({16}+\overline{16})$ and 
{10} representations of $SO(10)$. It is symmetric under exchange of 
rows and columns, reflecting the SVD that 
underlies the entire space of vacua.}
\end{minipage}
\end{figure}

The SVD was discovered in free fermionic constructions of the
heterotic--string in four dimensions by using the computerised 
classification tools that were developed for the analysis of
the spectrum of free fermionic vacua \cite{GKR, fknr, fkr, FKRSVD}.
The use of systematic computerised tools has been a hot pursuit over the
past years (for review and references see {\it e.g.} \cite{Ruehle}). 
The SVD was observed initially by simple counting \cite{FKRSVD},
which is illustrated in table \ref{svdcounting}, 
 where $s$, ${\bar s}$ and $v$ 
 are the total number of $16$, $\overline{16}$, and $10$
 representations of $SO(10)$ respectively. 
 It is easily seen by adding the corresponding numbers of models
 that the total number of vacua with
 two $16$,
 two $\overline{16}$, and
 one ${16}\oplus\overline{16}$, representations,
is the same as the total
number of models with two $10$ representations.
A more comprehensive depiction is illustrated in figure \ref{den}.
So far the SVD had been presented in terms of the free fermion
contructions. In the free fermionic classification method,
the set of boundary conditions is fixed and the variation
of the models is generated in terms of the one--loop Generalised--GSO (GGSO)
phases.
It is therefore aparent that in these constructions the
SVD arises due to exchange of GGSO projection coefficients,
which can be proven rigorously \cite{FKRSVD}.
Further insight into the SVD can be gained by using a bosonic
representation of the $Z_2\times Z_2$ orbifolds \cite{AFT, FFMT}. Since the
SVD operates plane by plane, it is sufficient to examine the
case with a single $Z_2$ twist. 

\begin{figure}[!]
	\centering
	\includegraphics[width=100mm]{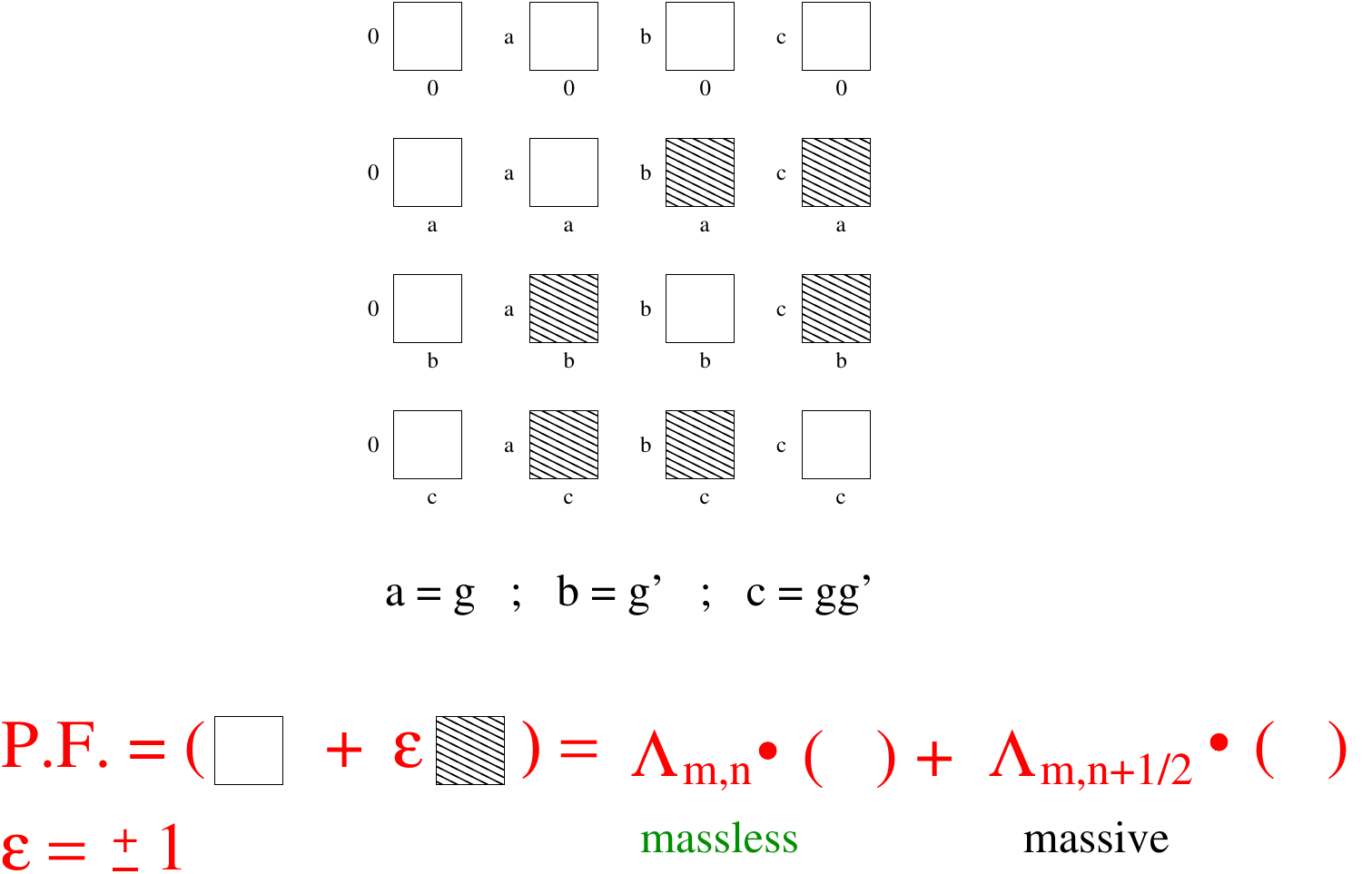}
	\caption{{
            The
            $Z_2\times Z_2^\prime$ partition function of the $g^\prime$--twist and $g$ Wilson line
            with discrete torsion $\epsilon=\pm1$. 
}
}
	\label{z2z2svd}
\end{figure}

Using the level one $SO(2n)$ characters,
\beqn
& & O_{2n} = {1\over 2} \left( {\theta_3^n \over \eta^n} +
{\theta_4^n \over \eta^n}\right) ~~~~~~\,,
~~~~~~~~~~~~
V_{2n} = {1\over 2} \left( {\theta_3^n \over \eta^n} -
{\theta_4^n \over \eta^n}\right) \,,
~~~~~~~~~~~~
\label{so2ncharaOV} \\
& & S_{2n} = {1\over 2} \left( {\theta_2^n \over \eta^n} +
i^{-n} {\theta_1^n \over \eta^n} \right) ~~\,,
~~~~~~~
C_{2n} = {1\over 2} \left( {\theta_2^n \over \eta^n} -
i^{-n} {\theta_1^n \over \eta^n} \right) \,,
\label{so2ncharaSC}
\eeqn
{where} 
\beq
{
\theta_3\equiv Z_f{0\choose0}~~~,~~~
  \theta_4\equiv Z_f{0\choose1}~~~},~~~
{
  \theta_2\equiv Z_f{1\choose0}~~~,~~~
  \theta_1\equiv Z_f{1\choose1}~,~~}\nonumber
\eeq
and $Z_f$ is the partition function of a complex worldsheet fermion. 
The partition function
of the $E_8\times E_8$  heterotic--string in four
dimensions is given by
\beq
{Z}_+ = (V_8 - S_8) \, 
\left( \sum_{m,n} \Lambda_{m,n}\right)^{\otimes 6}\, 
\left(\overline{ O} _{16} + \overline{ S}_{16} \right) 
\left(\overline{O} _{16} + \overline{ S}_{16} \right)\,,
\label{zplus}
\eeq
where for each $S_1$,
\beq
p_{\rm L,R}^i = {m_i \over R_i} \pm {n_i R_i \over \alpha '} \,
~~~~~~~~~
{\rm and}
~~~~~~~~~
\Lambda_{m,n} = {q^{{\alpha ' \over 4} 
p_{\rm L}^2} \, \bar q ^{{\alpha ' \over 4} 
p_{\rm R}^2} \over |\eta|^2}\,.
\nonumber
\eeq
A $Z_2\times Z_2^\prime:g\times g^\prime$ action on $Z_+$ is performed.
The first $Z_2$ couples a fermion number in the observable and hidden
sectors with 
a $Z_2$--shift in a compactified coordinate, and is given by
$
g: (-1)^{(F_{1}+F_2)}\delta.
$
Here the fermion numbers $F_{1,2}$ operate on the spinorial
representations of the observable and hidden $SO(16)$ groups as
$$
F_{1,2}:({\overline O}_{16}^{1,2},
             {\overline V}_{16}^{1,2},
             {\overline S}_{16}^{1,2},
             {\overline C}_{16}^{1,2})\longrightarrow~
            ({\overline O}_{16}^{1,2},
             {\overline V}_{16}^{1,2},
             -{\overline S}_{16}^{1,2},
             -{\overline C}_{16}^{1,2})
$$
and $\delta$ identifies points shifted by a $Z_2$ shift 
in the $X_9$ direction, {\it i.e.} 
$
\delta X_9 = X_9 +\pi R_9.~
$
The result of the shift is to insert a factor of $(-1)^m$ into the lattice 
sum in eq. (\ref{zplus}), {\it i.e.} 
$
\delta:\Lambda_{m,n}^9\longrightarrow(-1)^m\Lambda_{m,n}^9.
$
The second $Z_2$ is a twist of the internal coordinates given by 
\beq
{g^\prime}:(x_{4},x_{5},x_{6},x_7,x_8,x_9)
\longrightarrow
(-x_{4},-x_{5},-x_{6},-x_7,+x_8,+x_9). 
\label{z2twist}
\eeq
Alternatively, the first $Z_2$ action can be interpreted as a Wilson line
in $X_9$ \cite{FFMT},
$$
g~: (0^7,1|1, 0^7)  ~\rightarrow~
E_8\times E_8\rightarrow SO(16)\times SO(16).\label{wilsonline}
$$
The $Z_2$ twist in the internal space breaks $N=4\rightarrow N=2$ spacetime
supersymmetry and $E_8\rightarrow E_7\times SU(2)$, or
with the inclusion of the Wilson line 
$SO(16)\rightarrow SO(12)\times SO(4)$.
The orbifold partition function is
$${Z~=~
\left({Z_+\over{Z_g\times Z_{g^{\prime}}}}\right)~=~
\left[{{(1+g)}\over2}{{(1+g^\prime)}\over2}\right]~Z_+}.$$
The partition function contains an untwisted sector and three twisted sectors.
Its schematic form is shown in figure \ref{z2z2svd}.

The winding states in the sectors twisted by 
$g$ and $gg^\prime$ are shifted by $1/2$. Consequently, these sectors 
contain only massive states. The 
$g^\prime$ twisted sector produces massless matter states. 
The partition function has one discrete torsion $\epsilon=\pm1$
between the two modular orbits, and produces
massless states for zero lattice modes. 
The terms in the $g^\prime$ twisted sector 
contributing to the massless 
spectrum have the form
\beqn
& &     \Lambda_{p,q}
\left\{
 {1\over2}
\left( 
           \left\vert{{2\eta}\over\theta_4}\right\vert^4
         +
           \left\vert{{2\eta}\over\theta_3}\right\vert^4
\right)
\left[{
       P_\epsilon^+Q_s{\overline V}_{12}{\overline C}_4{\overline O}_{16}} +
   {P_\epsilon^-Q_s{\overline S}_{12}{\overline O}_4{\overline O}_{16} }
\right.
{\left.  \right] + }
\right. \nonumber\\ 
& &\nonumber\\ 
& &\left.
~~~~~~~~~~{1\over2}\left(    \left\vert{{2\eta}\over\theta_4}\right\vert^4
                      -
                         \left\vert{{2\eta}\over\theta_3}\right\vert^4\right)
\left[{
P_\epsilon^+Q_s
{\overline O}_{12}{\overline S}_4{\overline O}_{16}} \right.
{\left. \right] } 
\right\}~~~~~
+~~\hbox{massive}
 \label{masslessterminpf}
\eeqn
where 
\beq
P_\epsilon^+~=~\left({{1+\epsilon(-1)^m}\over2}\right)\Lambda_{m,n}~~~;~~~
P_\epsilon^-=\left({{1-\epsilon(-1)^m}\over2}\right)\Lambda_{m,n} 
\label{pepluspeminus}
\eeq

From the sign of the discrete torsion $\epsilon=\pm$ 
it is noted from  eq. (\ref{pepluspeminus}) that either the vectorial states, 
or the spinorial states, are massless.
It is readily seen from eq. (\ref{eplus}) and eq. (\ref{eminus})
that the choice $\epsilon=+1$ gives rise to 
massless momentum 
modes from the shifted lattice in $P_\epsilon^+$ whereas 
$P_\epsilon^-$ produces massive modes. Therefore, 
the vectorial character ${\overline V}_{12}$ in eq. (\ref{pepluspeminus})
gives rise to massless states, whereas the spinorial character
${\overline S}_{12}$ produces massive states.
Eq. (\ref{eminus}) shows that the choice $\epsilon=-1$ produces
exactly the opposite result. 
\beqn
{\epsilon~=~+1~~}&{\Rightarrow}&
{~~P^+_\epsilon~=~~~~~~~~~~~\Lambda_{2m,n}~~~
~~~~~~~~~~~~{ P^-_\epsilon~=~~~~~~~~
\Lambda_{2m+1,n}}~~~}\label{eplus}\\
{\epsilon~=~-1~~}&{\Rightarrow}&
{{ ~~P^+_\epsilon~=~~~~~~~~~~~\Lambda_{2m+1,n}}~~~
~~~~~~~~~P^-_\epsilon~=~~~~~~~~\Lambda_{2m,n}~~~}\label{eminus}
\eeqn
The spinor--vector duality is generated due to the exchange of the
discrete torsion
$\epsilon=+1\rightarrow \epsilon =-1$
in the $Z_2\times Z_2^\prime$ partition function.
This is similar to the case of mirror symmetry in the $Z_2\times Z_2$
orbifold of ref. \cite{vafawitten}, where the mirror symmetry
transformation results
from the exchange of the discrete torsion between the two $Z_2$ orbifold twists.

This particular example 
provides insight into the inner working of the SVD map. 
As shown in fig. \ref{den} and table \ref{svdcounting},
the SVD is manifested in the 
wider space of string vacua with $N=2$ and $N=1$ spacetime supersymmetry
\cite{fkr, FKRSVD, FKRNeq2}.
The analysis using the free fermionic formulation obscures the role of 
the geometrical moduli fields. In \cite{FKRSVD, FKRNeq2} it is shown
in terms of the GGSO projection 
coefficients of the one--loop partition function that the SVD always exists in this space of 
vacua.
The bosonic analysis in \cite{FFMT} reveals the role
of the moduli fields and demonstrates 
that the SVD arises due to an exchange of two Wilson lines. 
The SVD can then be interpreted to arise from the breaking of
the $N=2$ worldsheet supersymmetry on the bosonic side of the
heterotic--string. It was further shown that the map between
the dual vacua is induced in terms of a spectral flow
operator. At the enhanced self--dual point, the spectral flow operator
exchanges between the spinorial and vectorial components of the
representations of the enhanced symmetry group. In the vacua with broken
symmetry the spectral flow operator induces the map between the dual
Wilson lines and the dual vacua \cite{FKRNeq2, FFMT}. 
In ref. \cite{AFG} this picture was generalised to string vacua
with interacting internal CFTs \cite{AFG} that
utilise the Gepner construction \cite{gepner}.
The bosonic representation of the SVD is instrumental
for studying the imprint of the SVD in 
the effective field theory limit.

The details of the relation between the discrete torsion and the
Wilson line realisations of the SVD are discussed in ref. \cite{FFMT}.
It is sufficient here to 
realise that there are choices of the background moduli fields that
give rise to the spectra of the dual models.
The $Z_2$ twist action of the internal coordinates is given by
eq. (\ref{z2twist}), whereas the dual Wilson lines are given by 
\begin{align}\label{WL+}
		g = (0,0,0,0,0,1|0,0|1,0,0,0,0,0,0,0).
\end{align}
and 
\begin{align}\label{WL-}
	    g = (0,0,0,0,0,0|1,0|1,0,0,0,0,0,0,0),
\end{align}
and the map between the two is induced by the spectral flow 
operator \cite{FFMT}.
Relating the worldsheet symmetries to the properties
of the effective field theory limit of the string compactifications
is facilitated by 
using the bosonic data in the form of eqs. (\ref{WL+}) and (\ref{WL-}).
The interpretation of the worldsheet data in the effective field theory
limit is often obscured, as, for example, in the case of mirror symmetry.
For this purpose, the representation of the Spinor--Vector Duality in terms 
of the Wilson lines is particularly instrumental.

\section{Mirror symmetry}

Mirror symmetry was observed initially in worldsheet constructions of string
compactifications. Subsequently, the profound implications for complex 
geometrical manifolds that are used in the effective field theory
limit of the string compactifications
was understood.
Mirror symmetry facilitates the counting 
of intersections between sub--surfaces of the complex manifolds,
which are otherwise notoriously difficult to calculate.

The calculation is facilitated by the relation of the intersection curves to the calculation 
of the Yukawa couplings between the string states. Thus, the worldsheet constructions 
provide a useful tool to study the properties of the string vacua in the effective field theory limit. For brevity we can consider the mirror models in the
free fermionic formulation, though the mirror symmetry phenomena apply to
the whole space of string configurations.
The vacua with unbroken $SO(10)$ group are produced by a set of 
twelve basis vectors 
\begin{eqnarray}
v_1={\bf1}&=&\{\psi^\mu,\
\chi^{1,\dots,6},y^{1,\dots,6}, \omega^{1,\dots,6}~~~|
~~~\bar{y}^{1,\dots,6},\bar{\omega}^{1,\dots,6},
\bar{\eta}^{1,2,3},
\bar{\psi}^{1,\dots,5},\bar{\phi}^{1,\dots,8}\},\nonumber\\
v_2=S&=&\{\psi^\mu,\chi^{1,\dots,6}\},\nonumber\\
v_{3}=z_1&=&\{\bar{\phi}^{1,\dots,4}\},\nonumber\\
v_{4}=z_2&=&\{\bar{\phi}^{5,\dots,8}\},
\label{basis}\\
v_{4+i}=e_i&=&\{y^{i},\omega^{i}|\bar{y}^i,\bar{\omega}^i\}, \ i=1,\dots,6,
~~~~~~~~~~~~~~~~~~~~~N=4~~{\rm Vacua}
\nonumber\\
& & \nonumber\\
v_{11}=b_1&=&\{\chi^{34},\chi^{56},y^{34},y^{56}|\bar{y}^{34},
\bar{y}^{56},\bar{\eta}^1,\bar{\psi}^{1,\dots,5}\},
~~~~~~~~N=4\rightarrow N=2\nonumber\\
v_{12}=b_2&=&\{\chi^{12},\chi^{56},y^{12},y^{56}|\bar{y}^{12},
\bar{y}^{56},\bar{\eta}^2,\bar{\psi}^{1,\dots,5}\},
~~~~~~~~N=2\rightarrow N=1. \nonumber
\end{eqnarray}
The first ten vectors preserve $N=4$ spacetime supersymmetry 
and the last two are the $Z_2\times Z_2$ orbifold twists. The $e_i$
basis vectors correspond to shifts in the internal compactified
coordinates, whereas the $z_i$ basis vectors reduce the untwisted hidden sector
gauge group to $SO(8)\times SO(8)$. 
The third twisted sector of the $Z_2\times Z_2$ orbifold 
is obtained as the combination
$b_3= b_1+b_2+x$, where the $x$--sector is obtained from the
combination
\beq
x= {\bf1} +S + \sum_{i=1}^6 e_i +\sum_{k=1}^2 z_k =
\{{\bar\psi}^{1,\cdots, 5}, {\bar\eta}^{1,2,3}\}.
\label{xmap}
\eeq
The $x$--sector can give rise to additional massless spacetime vector
bosons in the observable sector. If these are not projected by the
GGSO projections, they enhance the $SO(10)$ gauge symmetry
to $E_6$, and the matter respresentations are in the $27$ and $\overline{27}$
of $E_6$. Mirror symmetry in the large space of free fermionic
vacua correspond to the exchange of the GGSO phase
\beq
\cc{b_1}{b_2} = +1 \rightarrow \cc{b_1}{b_2}= -1
\label{mirrorgsomap}
\eeq
that corresponds to the discrete torsion exchange of \cite{vafawitten}.
The effect in heterotic--string vacua with $E_6$ symmetry
is to flip the net chirality of the chiral representations,
which is counted by the Euler characteristic of the
internal manifolds
\beq
~~~\frac{\chi}{2}~~~=\# ({27}
- \overline{27})~~~\longrightarrow~~-\frac{\chi}{2}
\label{eulermap}
\eeq
In string compactifications with (2,2) worldsheet supersymmetry
there is a one--to--one correspondence between the chiral and anti--chiral
representations, and the complex structure and K\"ahler moduli of
the internal manifolds. 
In terms of the moduli fields of Narain toroidal compactfications,
the metric $G$, the antisymmetric tensor field $B$, and the Wilson
line moduli $W$, 
the mirror map (\ref{mirrorgsomap}, \ref{eulermap}) correspond to exchange
of the internal moduli, {\it i.e.} the metric field $G$ and the antisymmetric
tensor field $B$, which relate to the complex structure and K\"ahler moduli of
the complex Calabi--Yau manifolds. The mirror symmetry map
exchanges the complex structure and K\"ahler moduli of the
internal compactified manifold.

Mirror symmetry was first observed \cite{GreenePlesser, CLS}
in Gepner contructions \cite{gepner}
of heterotic--string compactifications. It was not foreseen by mathematicians
and was a complete surprise from their point of view
\cite{Katz, mirrorsymmetry}. Moreover,
it was shown to be instrumental in the field of enumerative geometry
in which the intersections between sub--surfaces of the complex
manifolds are counted \cite{COGP}. The observation of mirror symmetry
in the space of complex manifolds is a profound observation from
the purely mathematical point of view and led to important developments
in pure mathematics. 
In this respect we should note
that the string compactifications on Calabi--Yau entails the analysis
of the string vacua in their effective field theory limit. This is
an example where the symmetries of the ultra--violet complete string theory
have fundamental imprints on the effective string theory limit of
the string compactifications. 
The Yukawa couplings between massless states in the string spectrum
of the worldsheet vacua are given in terms of correlators among
vertex operators
$$
\langle{V_1^fV_2^fV_3^b\cdot\cdot\cdot\cdot V_N^b\rangle},$$
where the vertex operators are given by \cite{calculating}
\beqn
V^{f}_{(-{1\over2})}& = &{\rm e}^{(-{c\over2})}~{\cal L}^\ell~%
{\rm e}^{(i\alpha\chi_{_{12}})}~
{\rm e}^{(i\beta \chi_{_{34}})}~
{\rm e}^{(i\gamma\chi_{_{56}})}~\nonumber\\
%
& & \left(~{\prod_{j}}{\rm e}^{(iq_i\zeta_{j})}~
~\{\sigma's\}~
{\prod_{j}}{\rm e}^{(i{\bar q}_i{\bar\zeta}_{j})}~\right)\nonumber\\
%
& & {\rm e}^{(i{\bar\alpha}{\bar\eta}_{1})}~
{\rm e}^{(i{\bar\beta}{\bar\eta}_{2})}~
{\rm e}^{(i{\bar\gamma}{\bar\eta}_{3})}~
{\rm e}^{(iW_R\cdot{\bar J})}
{\rm e}^{(i{1\over2}KX)}~
{\rm e}^{({\rm i}{1\over2}K\cdot{\bar X})},\label{yukawas} 
\eeqn
and the different components entering eq. (\ref{yukawas})
are detailed in \cite{calculating}. The non--vanishing
correlators have to be invariant under all the string symmetries.
In the vacua with enhanced $E_6$ symmetry,
the couplings are
between three $27$ chiral representations of $E_6$, and
the mirror map implies that
$$~~~27\cdot27\cdot27~\longleftrightarrow~
\overline{27}\cdot\overline{27}\cdot\overline{27}$$
On the Calabi--Yau manifolds that describe the string vacua
in their effective field theory limit, the Yukawa couplings
correspond to intersection of curves. Thus, one finds imprints
of the worldsheet correlators in the geometrical data and use them
to analyse the properties of the corresponding manifolds.
Mirror symmetry proved its power in this domain by
providing a tool to analyse the geometry.
I should emphasize that it is not possible here to describe in 
to the field of mirror symmetry. Interested readers are referred
to Sheldon Katz book \cite{Katz} that provides a very lucid
introduction to the subject and the more in depth monograph
\cite{mirrorsymmetry}. Superficially, the analysis of the intersection
of the rational curves on Calabi--Yau manifolds is related to the
Yukawa couplings and therefore the calculation of the Yukawa
couplings, that are related to the Gromov--Witten invariants
provides a tool to analyse the geometrical data of the manifolds. The
message in the current paper is that the relation of spinor--vector
duality to mirror symmetry, {\it i.e.} both represents mappings under
transformations of the moduli parameters of
the Narain toroidal spaces, suggests that the spinor--vector duality may
have similar interesting mathematical implications in the Effective Field
Theory (EFT) limit. In this context it is noted that, similar to mirror
symmetry, the likely tool to be of use is the calculation of
Yukawa couplings among the string states, albeit in the case of SVD
the picture is complicated because it involves not only the internal
space, but also the vector bundles that correspond the the gauge degrees
of freedom on them.

\section{Spinor--vector duality in the EFT limit}

I will not delve in the discussion here into technical details
that can be found in the original literatue
\cite{FGNHH1,FGNHH2, FGNHH3, FGNHH4}. Rather, I will discuss the
spinor--vector duality in relation to mirror symmetry and articulate
future directions for research in light of this relation. Just as in the case
of mirror symmetry the SVD, which was first observed in worldsheet
constructions, may have profound implications for the mathematical
properties of the geometrical manifolds with vector bundles,
corresponding to the gauge degrees of freedom
of the heterotic--string. 
In refs. \cite{FGNHH1} and \cite{FGNHH2} the SVD was analysed in
the effective field theory limit of the string compactifications
in six and five dimensions, respectively. The analysis was
performed by starting with an orbifold model that exhibits SVD
and analysing the EFT limit on a smooth Calabi--Yau manifold
with vector bundle by smoothing the orbifold singularities
using well established technique in this context \cite{GNTW, GNHT}.
Ref. \cite{FGNHH1} analyses the SVD on $T^4/Z_2\times S^1$ in
five dimensions, 
by including a twist in the form of eq. (\ref{z2twist})
that acts on four internal coordinates
and a Wilson line in the form of eq. (\ref{WL+}) or (\ref{WL-})
on the additional circle. The subsequent step is to analyse the
resolution of this orbifold to a smooth $K3\times S^1$ by
using some massless states in the orbifold model to blow up
the singularities. We incorporate a discrete torsion in
the analysis of the orbifold model between the twist and
the Wilson line and its effect on the resulting massless
states. In the model that we analyse the states used for the
resolution transform under the $SO(10)$ GUT symmetry. This entails
that the GUT symmetry is broken by the resolutions. As the available
states for the resolution differ in the dual configurations
and transform under the observable gauge symmetry, 
the gauge
degrees of freedom are also different in the two cases. This is
unlike the situation in some of the free fermionic SVD models
\cite{FKRSVD, FKRNeq2} that contain twisted hidden sector states that
may be used to resolve the singularities without affecting the
gauge symmetry. Since the role of the discrete torsion in the effective
field theory smooth limit is obscured, we make an educated guess on
the resolved manifold for the orbifold that includes the discrete
torsion. The case without torsion is fully well defined on the
resolved manifold, but the case with torsion introduces some
subtleties that are discussed in detail in ref. \cite{FGNHH1}. 
The short summary is that the smooth geometries do exhibit a
spinor--vector duality--like phenomenon, but that due to the
different spectra available for the resolutions on the
dual configurations, the gauge symmetries differ on the resolved
manifolds. This phenomenon is expected to be generic
in the resolved limit because of the different states
available for the resolution, {\it e.g.} the spinorials in
one case and the vectorials in the other, in the example discussed
in section \ref{svd}.
In ref. \cite{FGNHH2} the spinor--vector duality was studied
in six dimensions. It is found that the spinor--vector duality also
operates in these case, albeit the vacua are self--dual under
the spinor--vector dulaity map, and satisfy the general anomaly
consistence condition on the number of vector and spinor representations
of any $SO(2N)$ unbroken subgroup in the string vacuum
\beq
N_V=2^{N-5}N_S+2N-8. \label{anomalycond}
\eeq
The analysis of spinor--vector duality in smooth $Z_2\times Z_2$
orbifolds in four dimensions is complicated
due to the large number of possible resolutions. The $T^6/Z_2\times Z_2$
orbifold has 64 $C^3/Z_2\times Z_2$ singularities, where $Z_2$--fixed
tori intersect. All the singularities have to be resolved to produce
a smooth manifold. Each singularity can be resolved in four
topologically distinct ways \cite{FGNHH3}, resulting in
$4^{64}$ a priori disctinct possibilities. The symmetry structure
of the $Z_2\times Z_2$ orbifold can be used to reduce this number
but still leaves a large number, of order $10^{33}$ of distinct
configurations. Many of the physical properties of the effective
field theory limits of the resolved geometries, like the spectra
of massless states and the interactions between them, depend on
the chosen resolution, and hinder the extraction of generic
properties of the resolved $Z_2\times Z_2$ orbifolds. In ref.
\cite{FGNHH3} a formalism was developed that allows computations
of any choice of the resolution, which opens the way to extract
some properties of the resolved $T^6/Z_2\times Z_2$ that are independent
of the choice of the resolution, and therefore hold for any such
choice. The analysis of the spinor--vector duality in four dimensions
to this date is still outstanding.

Another tool in the analysis of the effective field theory limit of worldsheet
string models is Gauged Linear Sigma Models (GLSM) \cite{GLSM},
which provide a tool to interpolate between the singular orbifold
constructions and their resolved smooth geometries.
Some of the properties of the worldsheet string constructions that
do not have a direct analogue in the smooth geometries can therefore
be studied by using the GLSMs. An example of this is
the discrete torsion that appears in the worldsheet string vacua between
the different modular orbits in the string partition function and
has no direct analogue in the smooth geometries that underlie the
effective field theory limit. In ref. \cite{FGNHH4} we used the GLSM
to shed light on what becomes of the discrete torsion in the resolution
of non--compact $C^3/Z_2\times Z_2$ and the compact
$T^3/Z_2\times Z_2$ orbifolds. The GLSMs associated with the non–compact
orbifold with or without
torsion are to a large degree equivalent:
only when expressed in the same superfield basis, a field
redefinition anomaly arises among them, which in the orbifold limit
reproduces the discrete torsion phases.
The GLSMs associated with the torsional compact orbifold suffers
from mixed gauge anomalies, which need to
be cancelled by appropriate logarithmic superfield dependent
Fayet--Iliopoulos terms on the worldsheet, signalling
$H$–-flux due to NS5--branes supported at the exceptional cycles.

\section{Questions for future explorations}

As discussed above mirror symmetry is the key example of the
relation between worldsheet string constructions and their Effective Field
Theory (EFT) limit on smooth geometries. Mirror symmetry, that was discovered
in worldsheet string constructions, relates couplings in the
dual string vacua, which in the EFT limit on complex geometries
correspond to intersections of rational curves on Calabi--Yau manifolds.
Mirror symmetry proved to be instrumental in counting the numer of
such intersections, {\it i.e.} it proved to be a useful tool in the
purely mathematical field of enumerative geometry.
Dedicated tools, such as the Gromov--Witten invariants
were developed for that purpose.

The Spinor--Vector Duality (SVD) is an extension of mirror symmetry
in the sense described in section \ref{svd}. As such it is natural to
ask whether the SVD can be instrumental as a tool to
explore the properties of algebraic complex curves
with vector bundles on them and in particular to
explore the effective field theory limits of
worldsheet string constructions.
We can pursue that in the first instance by analysing the
correlators between the string states in the spinor--vector
dual vacua, and seek to define the analogues of the Gromov--Witten
invariants. The SVD provides a tool to study the complex Calabi--Yau
manifolds with vector bundles, which correspond to the gauge
degrees of freedom of the heterotic--string. The SVD thus provides a tool
to study the moduli spaces of (2,0) string compactifications. In this
respect we can ask whether it is complete?, {\it i.e.} does the SVD
constrain the viable effective field theory limits of quantum gravity
models that are compatible with the ultra--violet complete heterotic--string
theory. We can pose a ``Swampland'' conjecture \cite{spwsp}: ``Every EFT
(2,0) heterotic--string compactification which has an ultra--violet
complete embedding in string theory is connected to a (2,2)
heterotic--string compactification by an orbifold or
by continuous interpolation''. If it is not, then it is necessarily
in the ``Swampland''. The motivation to pose this cojecture stems
from the question whether the symmetries of the string worldsheet
formalism are complete. We can view this in analogy with the
celebrated $T$--dualities and mirror symmetry, where similarly, we may
question whether a mirror manifold should always exist and whether
$T$--duality represent a complete symmetry of string theory, {\it i.e.}
any string compactification must admit a symmetry that can be interpreted
as $T$--duality and can be connected to the self--dual point.
That is: Does $T$--duality provide a complete characterisation of
string theories of quantum gravity or is it merely a property of
string compactification on tori? In this respect, we can note that
the interpretation of $T$--duality as phase--space duality
(see {\it e.g.} \cite{sdvs} and references therein) 
may provide a generalisation that extends its realisation
beyond the toroidal geometry. The Spinor--Vector Duality
extends $T$--duality in the sense discussed in section \ref{svd}, {\it i.e.}
by including transformations induced due to exchange of Wilson line moduli
rather than the moduli of the internal compactified torus. The self--dual
point under the SVD is the enhanced $E_6$ symmetry point with $(2,2)$
worldsheet supersymmetry, or enhanced $E_7$ symmetry in the case of
a single $Z_2$ twist with $N=2$ spacetime supersymmetry. The transformation
between the dual Wilson lines is continuous in the later
case and discrete in the former. In the discrete case, the moduli
that enable the continuous interpolation in the $E_7$ case, are simply
projected out from the spectrum by the second $Z_2$ twist, and the
map between the dual vacua is discrete. In both cases the models are
connected to the self--dual enhanced symmetry point by either
a continuous interpolation or by an orbifold. This is similar
to the case of $T$--duality in which the continuous interpolation can
be nullified by asymmetric boundary condition assignments that project
some or all of the internal torus moduli \cite{moduli, FGNP}. 
We can conjecture that:

\noindent
    {\it SVD--conjecture:
      Every EFT (2, 0) heterotic–string compactification has to be connected to
a (2,2) heterotic--string compactification by an orbifold or by continuous
interpolation. Otherwise, it is in the swampland, i.e. it does not
have an ultra--violet completion in string theory.}
\indent

The approach articulated here therefore presents a top--down approach
to the Swampland--program \cite{SVDandSL}. The aim is to explore how
the symmetries of the ultra--violet complete string theories, that are
defined in terms of the worldsheet constructions, constrain the
effective field theory limits of these theories. The
``Swampland''--program approach aims to explore which effective
field theories of quantum gravity have an embedding in an ultra--violet
complete string theory of quantum gravity
(for review and references see {\it e.g.} \cite{swampland}) and can
be viewed as a bottom--up approach to the construction of
consistent theories of quantum gravity.

The SVD conjecture therefore provides a demarcation line between (2,0)
effective field theories that do, and do--not,
possess an ultra--violet complete embedding in string theory.
We can envision that there exist many
(2,0) EFTs that do not satisfy the SVD--conjecture
and those will be in the ``Swampland'', whereas those that do
satisfy the conjecture have an ultra--violet complete embedding
in string theory.
It should be noted that the SVD and $T$--duality are
merely two examples of the symmetry structure of $(2,0)$ string vacua,
and it is
anticipated that a much larger symmetry structure underlies them
\cite{panosmoon}. The proposition of the SVD--conjecture
as posed above is a physicist's proposition and making it
a proper mathematical statement is warranted. Likewise, we can ask, what
are the tools that can substantiate this statement, and it seems
that the GLSMs might provide such a tool.

\section{Conclusions}

Our understanding of fundamental physics reached a juncture in which
the mathematical description of all sub--atomic data is well accounted
for by the Standard Model (SM) of particle physics, whereas observations at the
celestial, galactic and cosmological scales are well accounted
for by Einstein's general relativity. Yet the two theories are
fundamentally incompatible. The Standard Model gives rise to
a large vacuum energy, whereas observations using Einstein's general
relativity are compatible with a much smaller vacuum energy. Furthermore,
using the QFT framework that is used in the Standard Model to
calculate quantum gravitational effects is plagued with infinities and
therefore inconsistent. Yet the synthesis of the Standard Model with
gravity is inevitable. The sub--atomic observational data indicates
that further basic insight into the Standard Model parameters,
{\it e.g.} in its flavour sector, 
can only be gained by synthesising it with gravity.

String theory is a self--consistent theoretical framework that
accommodates perturbative quantum gravity with all the
ingredients that make up the Standard Model. Detailed
phenomenological models can be constructed that reproduce
the structure of the SM, and enable the development
of a phenomenological approach to quantum gravity. The characterisation
of string theory in the literature is often misleading.
It is often called the Theory of Everyting and is therefore
poorly portrayed as a final step in our
understanding of fundamental physics.
First of all we do not know what everything is. String theory
is not a final step, but rather the relevant question is what and whether
any of the ingredients of string theory are relevant in the real
physical world. We can relate to the QFTs that underlie
the Standard Model as point particle theories, whereas
string theory is a string particle theory. The concept of a
particle with specified properties is well defined in both.
They are facets of the same object. One is compatible with quantum
gravity and one is not. Likewise in the thriving field
of amplitudes the calculational methods are interchangeable.

String theory is also used as a tool to explore the fundamental
mathematical structures that underlie the theory and their properties.
The most celebrated example among those is that of mirror
symmetry that was first observed in worldsheet constructions of
string compactifications and its profound implications for complex
manifolds and enumerative geometry were subsequently understood.
Spinor--Vector Duality (SVD) is an extension of mirror symmetry that
extends the duality map to include transformation of Wilson--line
moduli. The most general set of symmetries in toroidal orbifolds hence
act on the internal and Wilson line moduli as $(G,B,W)\rightarrow
({\tilde G}, {\tilde B}, {\tilde W})$, where $G$, $B$ and $W$
are the metric, anti--symemtric tensor and Wilson--line moduli
fields of the Narain moduli space.
Similarly to mirror symmetry the SVD was first noted
in worldsheet formulations of heterotic--string theories
and may have profound implications for the complex algebraic
curves with vector bundles that correspond to the
EFT limits of the worldsheet constructions. The SVD may provide
a demarcation line between the (2,0) quanum gravity EFTs that have an
ultra--violet complete embedding in string theory and those that do not.

The SVD is therefore of pure mathematical interest. We should note,
however, the use of SVD in the contruction of string derived
$Z^\prime$--model that may remain light down to low scales \cite{FRZp}
with implications for physics at the LHC \cite{FGZp}. 

\section*{Acknowledgements}

I would like to thank the Kavli Institute for Theoretical Physics,
the CERN theory division, and the Department of Particle Physics and
Astrophysics at the Weizmann Institute for hospitality.

\funding{
This research was supported in part by grant NSF PHY-2309135 to the
Kavli Institute for Theoretical Physics (KITP).}

\dataavailability{No new data were created or analyzed in this study. Data sharing is not applicable to this article}.

\conflictsofinterest{The authors declare no conflicts of interest.}

\reftitle{References}


\bibliography{MSSVD}


\PublishersNote{}
\end{document}